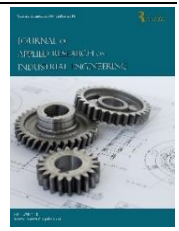

Paper Type: Research Paper

# Neural Network Based Human Reliability Analysis Method in Production Systems

**Rasoul Jamshidi**[1],*  , **Mohammad Ebrahim Sadeghi**[2]

[1] Department Industrial of Engineering, School of Engineering, Damghan University, Damghan, Iran; r.jamshidi@du.ac.ir;
[2] Department of Industrial Management, Faculty of Management, University of Tehran, Tehran, Iran; sadeghi.m.e@gmail.com.

**Citation:**

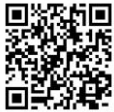

## Abstract

Nowadays, many accidents, malfunctions, and quality defects are happening in production systems due to Human Errors Probability (HEP). Human Reliability Analysis (HRA) methods have been proposed to measure the HEP based on Performance Shaping Factors (PSFs), but these methods do not have a procedure to select the effective PSFs and consider the PSFs dependency. In this paper, we propose an Artificial Neural Network based Human Reliability Analysis (ANNHRA) in cooperation with Response Surface Method (RSM). This framework uses the advantage Systematic Human Error Reduction and Prediction Approach (SHERPA) method to quantify the PSFs and the ANN and RSM to consider the PSFs dependency and select the most effective PSFs. This framework decreases the time and cost and increases the accuracy of HRA. The proposed framework has been applied to a real case and the provided results show that human reliability can be calculated more effectively using ANNHRA framework.

## 1 | Introduction

In each production system, human error is the main source of concern for managers. Although many production systems have been automated by using digital machines and equipment, human error has not been eliminated considerably and a high proportion of errors in many industries are derived from humans. For example, in chemical production systems 60-90% of error happens because of human failure [1].

Human errors can be divided into two categories according to their impact on production systems. The low and high impact categories. Low impact leads to produce poor quality and defective production [2]. But in critical industries such as nuclear or chemical industries, human errors in the high impact category can cause a fatal failure and impose a high risk to production systems [3]. Human errors also decrease productivity and increase undesirable costs such as idle cost, backorder cost, and quality cost.

Corresponding Author: r.jamshidi@du.ac.ir
10.22105/jarie.2021.277071.1274

Many researchers investigated human error and attempted to quantify the Human Errors Probability (HEP) as a function of the human type, work type and environmental factors [4]. These attempts lead to propose Human Reliability Analysis (HRA) methods by researchers.

The HRA methods were developed with the quantitative methods in the first generation and the qualitative methods in the second one, and in recent year dynamic HRA attracted the researchers' attention as the third generation [5]-[6]. The first generation includes many methods such as Technique for Human Error Rate Prediction (THERP) [7], Human Cognition Reliability (HCR)[8]. The first generation methods use a simple "fits" or "doesn't fit" dichotomy to match the error scenario to error identification and quantification. But second generation methods such as Cognitive Reliability and Error Analysis Method (CREAM)[9] and [10] use a theory-based error taxonomy and complex match of error scenarios to error identification and quantification. In these methods Performance Shaping Factors (PSFs) with multiple levels of assignment utilized to indicate degrees of degraded or enhance performance relative to nominal.

In recent years, dynamic HRM has become an interesting issue for study since many parameters influence human reliability and the effect of these parameters on human reliability cannot be calculated by the exact algorithm. Chang and Mosleh [11]-[15] room operating crew in nuclear power plants to use this method in proposed the Information Decision and Action in Crew (IDAC) context for HRA. The model was developed to predict the responses of the control Probabilistic Risk Assessments (PRA). Trucco and Leva [16] developed a new probabilistic cognitive simulator (PROCOS) to obtain the errors of human in operational systems, they used the quantification susceptibilities of the first-generation HRA with a cognitive evaluation for an operator. Kim et al.[17] proposed a scheme to classify the erroneous behaviors identified by simulator data. Pasquale et al. [18] proposed the Simulator for Human Error Probability Analysis (SHERPA), and utilized the advantages of the simulation technique and the traditional HRA methods to model human behavior and obtain the error probability for a specific scenario in production systems. In these models, the effect of PSFs has been simulated and the results are used to predict the HEP [18]. Pandya et al. [19] proposed a methodology to support a systematic and traceable process to develop the generic task type–performance-influencing factor structure, to ease the review of the HRA process. Pandya et al. [20] proposed a model for HRA quantification based on expert judgment aggregation. Laumann investigated the quality criteria for qualitative HRA method to make the HRA results more accurate [21].

Simulation of the PSFs effect on HEP requires the occurrence probability of each PSFs and the manner of PSFs effect on HEP. In many real situations, these data are not available or accurate. Also, most HRA methods have been developed in a specific context, such as nuclear power plants and fewer methods have been proposed for production systems. Also, some HRA methods such as HEART (Human Error Assessment and Reduction Technique) and SHERPA proposed some PSFs and calculate the HEP base on these factors, lack of attention to the factors dependency is one of the shortcomings of these methods that causes error in HEP calculation. All PSFs should be considered together to investigate their effect on human error. Another issue is that not all PSFs need to be used and the most important ones should be considered to reduce the cost and time of the HRA. In this paper, we propose a method to eliminate the mentioned defects by Artificial Neural Networks (ANN) and Response Surface Method (RSM), The ANN considers the PSFs dependency and RSM selects the most important PSFs. This method reduces the cost, and time in HEP calculation and makes results closer to reality. This method uses the neural network to calculate the HEP based on several environmental, work (duration, type), and human (age, sex) factors and their dependency, and RSM to investigate the effect of each PSF on human error and eliminated PSFs with insignificant effect. Since we consider factors dependency, the results are more accurate and reliable.

The rest of the paper is organized as follows: Section 2 presents the research methodology and describes the HEP calculation method, PSFs and ANN. Section 3 is the finding and discussion section that describes the ANNHRA framework and proposes the framework procedure for a real case, and finally, Section 4 concludes the paper.

## 2 | Research Methodology

In this section, the methods used in this paper such as HEP calculation method, PSFs, ANN and RSM are presented.

### 2.1 | Human Error Probability Calculation

According to The THERP method, HRA aims to find the contribution of human reliability to the system reliability, that is to say, the aim is to predict human error probability and assess the total unreliability of human–machine systems likely to be caused by a human in association with equipment, machines, procedures, and human characteristics which influence the production system [22]. The first step in the HRA is error identification. In this step, all probable errors should be identified with their consequences. The second step is calculating the occurrence probability of each identified error and the final step is the reduction of error probability. Kirwan [23] proposed that the HEP can be calculated as follows:

$$\text{HEP}_{nominal} = \frac{\text{OEN}}{\text{PEN}}. \qquad (1)$$

The OEN is the number of occurred errors and PEN is the number of potential errors. For example, if the opportunity for a specific error is 20 times and the error occurs 10 times the HEP is equal to 50%. It should be noted that data gathering to use this formula is not simple, some researchers presented that HEP is derived from four Contextual Control Modes (CoCoMs), scrambled, opportunistic, tactical, and strategic in the first generation and several PSFs such as stress and complexity, but dynamic HRA presented that HEP is a result of human performance factor relations and dependencies such as work type and work time [24].

### 2.2 | Performance Shaping Factors

PSF was advocated by Swain [25] first time and is usually treated as "the regulation item for the introduction of the error rate" or "the providing items for the prediction of human error". PSFs are the aspects of human behavior and the context that can impact human resource performance, these factors were viewed in terms of the effects, they might exert on human performance such as work efficiency and system reliability. Many PSFs and categories have been proposed by researchers for different systems such as nuclear or power plants [26], [27] and [28].

In practice, the number of PSFs that are included in HRA methods lies between these 1 to 60 PSFs. For example, the SPAR-H method [22], which is widely used in the US nuclear industry, includes eight PSFs. The internationally widely used CREAM [29] uses nine PSFs. Boring studied the important PSFs and proposed 8 PSFs that are considered in common HRA methods [30]. These PSFs are as follows:

– *Available time,*
– *Stress,*
– *Complexity,*
– *Experience and training,*
– *Procedures,*
– *Ergonomics,*
– *Fitness for duty,*
– *Work process.*

PSFs may have a negative impact or a positive impact on human error. When the influencing factor represents a positive impact, it corresponds to a value less than one; which is used to decrease the HEP value. Also, the PSF represents a negative impact, it corresponds to a value greater than one and leads to decreases the HEP. The total impact of PSFs is calculated using *Eq. (2)*.

$$\text{Total PSFs Impact} = \text{IPSF}_1 * \text{IPSF}_2 * \ldots * \text{IPSF}_8. \tag{2}$$

*Eq. (2)* shows that the total impact of PSFs is the multiplication of each PSF impact (IPSF). The HEP formulation has been shown by *Eq. (3)*.

$$\text{HEP}_{\text{composite}} = \frac{\text{HEP}_{\text{nominal}} \cdot \text{PSF}_{\text{total}}}{\text{HEP}_{\text{nominal}} \cdot (\text{PSF}_{\text{total}} - 1) + 1}. \tag{3}$$

Although *Eq. (3)* is used to obtain the human error probability considering the PSFs in some mentioned methods. In this paper, we investigate the impact of PSFs on HEP separately, the PSFs impact on HEP are evaluated by ANN. That is to say, in the proposed method the value of HEPcomposite is predicted by ANN and there is no need to calculate the HEPnominal in the first step.

### 2.3 | Artificial Neural Network

ANN are well known since they can process a huge amount of information through an interconnected network with several nodes in many layers [31]. The information flow is related to the network architecture which mimics the one present in the neurons of the human body. Since ANN is an imitation of nature and the human body, it includes several steps such as recognition, verification, optimization, and prediction. The ANN aims to understand the effect of parameters (inputs) on a result (output) in different systems. After selecting the input parameters, the learning process starts with the training and testing steps. Also, it can be done after standardizing and eliminating the outliers of the input parameters [32]. The schema of ANN has been shown in *Fig. 1*.

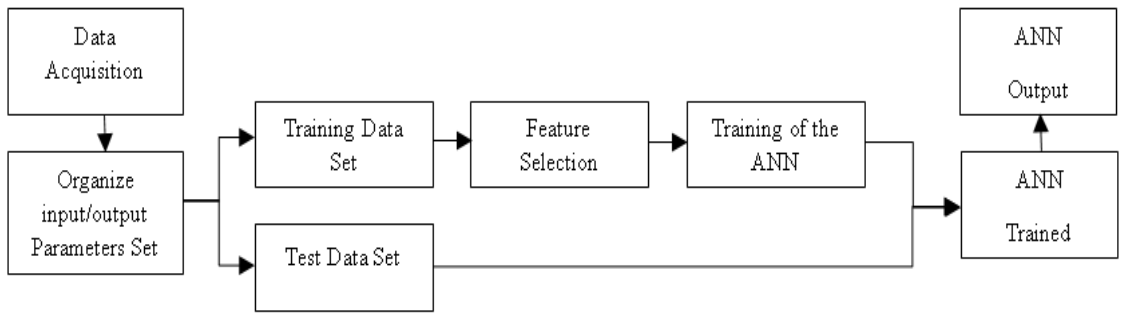

**Fig. 1. The schema of ANN.**

In the learning process, different networks can be selected and examined for obtaining their performance by altering the layers and nodes number, the transfer functions' shape, and the learning algorithm. To evaluate the accuracy of different networks, *Eq. (4)* minimizes the overall error between the targets and calculated values.

$$R^2 = 1 - \frac{\sum_{i=1}^{N} (y_i' - y_i)^2}{\sum_{i=1}^{N} (y_i' - \sum_{i=1}^{N} y_i'/N)^2}. \tag{4}$$

In *Eq. (4)* $y_i'$ and $y_i$ are the actual and estimated output values, and N is the number of input data points.

In this paper, we use the ANN to obtain the effect of PSFs on the HEP. In the prior researches, some coefficients have been proposed for each PSF that show the effect of PSF on the HEP. For example, the SPAR-H method proposed a table for PSFs value in different situations and the systems. *Table 1* shows the weight for different manners of available time as a PSF.

Table 1. The PSF weight for available time.

| PSF level (Available Time) | Multipliers Action | Multipliers Diagnosis |
|---|---|---|
| Inadequate Time | P (failure) =1 | P (failure) =1 |
| Time available = time required/Barely adequate time | 10 | 10 |
| Nominal time | 1 | 1 |
| Time available > 5 x time required Extra time | 0.1 | 0.1 |
| Time available > 50 x time required Expansive time | 0.01 | 0.01 |
| Insufficient information | Nominal time | Nominal time |

As could be seen in *Table 1* the system and environment have no effect on PSF weight, but the weight of the PSF level should be varied based on system type and environment conditions. Considering this fact, all PSF levels have some error and multiplication of this error leads to a significant error in total HEP.

In this paper, we provide the actual effect of the PSFs level by using the ANN method. Using this method can make us sure that the total HEP value has not a significant error and is reliable. Selecting the important PSFs and the environmental conditions depends on the experts' opinion and in experts' opinion we confront a specific level of error and inconsistency. Using ANN can reduce this error and predict more accurate HEP value based on PSFs in different work environments.

## 2.4 | RSM

The RSM has been proposed by box and his collaborators [33]. This method was derived from the multi-dimensional graph to assess the fitness of the mathematical model. RSM consists of a group of techniques that aim to fit an empirical model based on the experimental data, to achieve this goal, linear or square polynomial functions are employed to describe the system studied and, to optimize the experimental conditions to propose the optimal configuration of input parameters [34]. RSM can be summarized in six steps. The first step is the selection of independent inputs that influence the results. The second step is designing an experimental matrix and carrying out the experiments according to the selected experimental matrix. The third step is fitting a function to obtain the mathematic–statistical treatment of experimental data. The fitness evaluation of the obtained model is the fourth step. Verification of the necessity and possibility of performing a displacement in direction for the optimal region is the fifth step. Finally providing the optimum value for each input is the sixth step. The application of RSM in the proposed framework is described in the next section.

## 3 | Finding & Discussion

In this section, we propose the ANNHRA framework and its procedure to calculate the human error and reliability, and examine the effectiveness of the proposed framework using a set of data from a real case. The provided results verify the framework effectiveness in human error and human reliability calculation.

### 3.1 | ANNHRA Framework

he proposed framework to assess human reliability is shown in *Fig. 2*. As could be seen the framework has two subsystems, the first is ANN that evaluates the effect of PSFs and the second is RSM which identifies the most effective PSFs in different systems. This framework will be repeated until the results converge. This framework proposed the HEP value and provide the most effective PSFs in different systems that lead to reduce the time and cost in data gathering and increase the accuracy of HEP value.

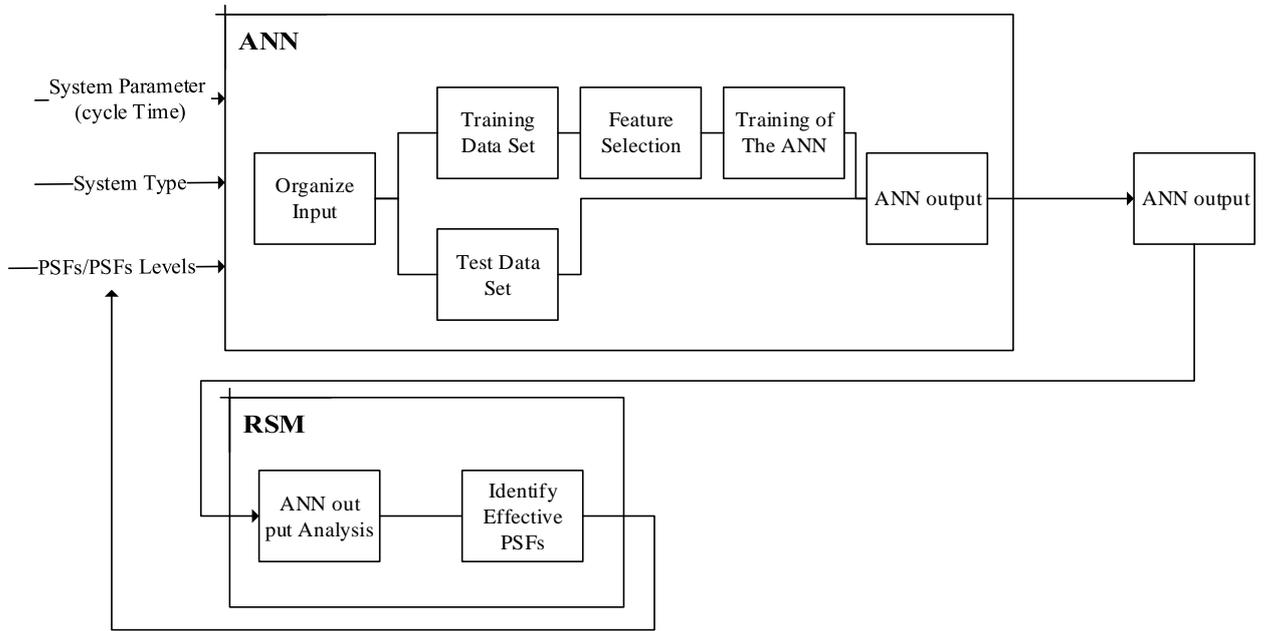

**Fig. 2. The proposed framework to assess the human reliability**.

## 3.2 | Framework Procedure

To illustrate the procedure of the proposed framework we investigated a real case. In this case, a lathing workshop with three human resources and 8 PSFs have been considered. The aim is to find the most important PSFs and their dependencies for calculating the HEP to reduce the cost of gathering information and sampling in the production system. First of all, we used some samples to provide HEP and ANN to predict HEP for other sets of PSFs. We considered several works with different conditions and calculated the related PSFs value according to the works and their implementation conditions. The PSFs multipliers like what was mentioned in *Table 1* can be found in [35] and [36]. To calculate the HEP, we implemented a set of works by a set of workers and calculated the failure probability by doing works repeatedly (at least 100 times for each work). *Table 2* shows the HEP and PSFs for each work.

**Table 2. Instances detail PSFs.**

| | Available Time | Stress | Complexity | Experience And Training | Procedures | Ergonomics | Fitness For Duty | Work Process | HEP |
|---|---|---|---|---|---|---|---|---|---|
| Ins 1 | 0.1 | 2 | 5 | 3 | 20 | 0.5 | 5 | 0.5 | 0.155 |
| Ins2 | 10 | 2 | 2 | 0.5 | 1 | 10 | 1 | 0.5 | 0.13 |
| Ins 3 | 10 | 1 | 2 | 1 | 50 | 1 | 5 | 5 | 0.15 |
| Ins 4 | 0.1 | 1 | 1 | 1 | 5 | 0.5 | 1 | 0.5 | 0.04 |
| Ins 5 | 10 | 2 | 5 | 3 | 50 | 1 | 5 | 5 | 0.2 |
| Ins 6 | 0.1 | 2 | 5 | 0.5 | 1 | 1 | 5 | 1 | 0.09 |
| Ins 7 | 0.01 | 1 | 1 | 0.5 | 1 | 1 | 1 | 5 | 0.03 |
| Ins 8 | 1 | 5 | 2 | 1 | 1 | 1 | 1 | 5 | 0.1 |
| Ins 9 | 0.01 | 5 | 2 | 0.5 | 1 | 10 | 5 | 5 | 0.165 |
| Ins 10 | 10 | 2 | 1 | 3 | 1 | 10 | 5 | 5 | 0.18 |
| Ins 11 | 1 | 1 | 5 | 0.5 | 1 | 10 | 1 | 5 | 0.11 |
| Ins 12 | 0.1 | 5 | 5 | 3 | 50 | 10 | 1 | 1 | 0.19 |
| Ins 13 | 0.01 | 5 | 5 | 0.5 | 20 | 10 | 5 | 0.5 | 0.17 |
| Ins 14 | 0.1 | 5 | 2 | 3 | 5 | 0.5 | 5 | 1 | 0.15 |
| Ins 15 | 0.1 | 5 | 2 | 3 | 5 | 0.5 | 5 | 1 | 0.16 |

As could be seen, several combinations of PSFs and their effects on the HEP have been calculated according to work and the workers' conditions. By using ANN, the framework aims to calculate the HEP and provide a function to determine the HEP according to PSFs.

In the first step, the PSFs value should be normalized to [0,1]. In this case, a simple normalization scheme is adopted as follows:

$$\text{PSF}_i^{j\ norm} = \frac{\text{PSF}_i^j}{\text{PSF}_{max}^j}. \tag{5}$$

The normalized PSFs and their related HEP are fed into the network shown in *Fig. 3* for training. There are many algorithms for ANN training. These algorithms usually search ANN parameters that minimize the deviation of predicted values from the measured values.

The proposed ANN has three layers: the input layer, the hidden layer, and the output layer. It has 8 inputs (number of PSFs) with one output (The HEP value). To overcome the over fitting problem, the hidden node number should not be very large. With some practical guidelines, this number is selected to be the same as the number of input nodes.

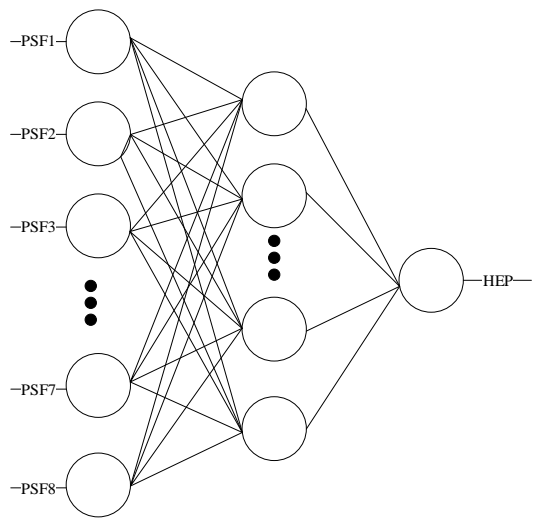

**Fig. 3. The ANN layers schema.**

The results of the proposed ANN for historical data are shown in *Table 3*. Since the ANN is initialized randomly before training, several replicated runs are usually done with different initialization, and the average is used as the estimated HEP. To assess the performance of the proposed ANN the Square Error (SE) is calculated for each instance. According to SE value, the Mean Square Error (MSE) is calculated to evaluate the fitness of the ANN function. The MSE is equal to 5.24E-04 for *Table1*.

To make it clearer, the HEP and Estimated HEP are illustrated in *Fig. 4*. As could be seen the difference between HEP and Estimated HEP is large in some instances. To eliminate this tolerance the RSM should be used to refine the PSFs and select the most effective PSFs.

Table 3. The results of ANN.

|  | HEP | Estimated HEP (ANN) | SE |
|---|---|---|---|
| **Ins 1** | 0.155 | 0.134 | 4.41E-04 |
| **Ins2** | 0.132 | 0.161 | 8.41E-04 |
| **Ins 3** | 0.151 | 0.162 | 1.21E-04 |
| **Ins 4** | 0.046 | 0.064 | 3.24E-04 |
| **Ins 5** | 0.223 | 0.263 | 1.60E-03 |
| **Ins 6** | 0.098 | 0.0851 | 1.66E-04 |
| **Ins 7** | 0.032 | 0.046 | 1.96E-04 |
| **Ins 8** | 0.136 | 0.112 | 5.76E-04 |
| **Ins 9** | 0.165 | 0.175 | 1.00E-04 |
| **Ins 10** | 0.182 | 0.154 | 7.84E-04 |
| **Ins 11** | 0.112 | 0.127 | 2.25E-04 |
| **Ins 12** | 0.193 | 0.175 | 3.24E-04 |
| **Ins 13** | 0.172 | 0.196 | 5.76E-04 |
| **Ins 14** | 0.153 | 0.186 | 1.09E-03 |
| **Ins 15** | 0.164 | 0.142 | 4.84E-04 |

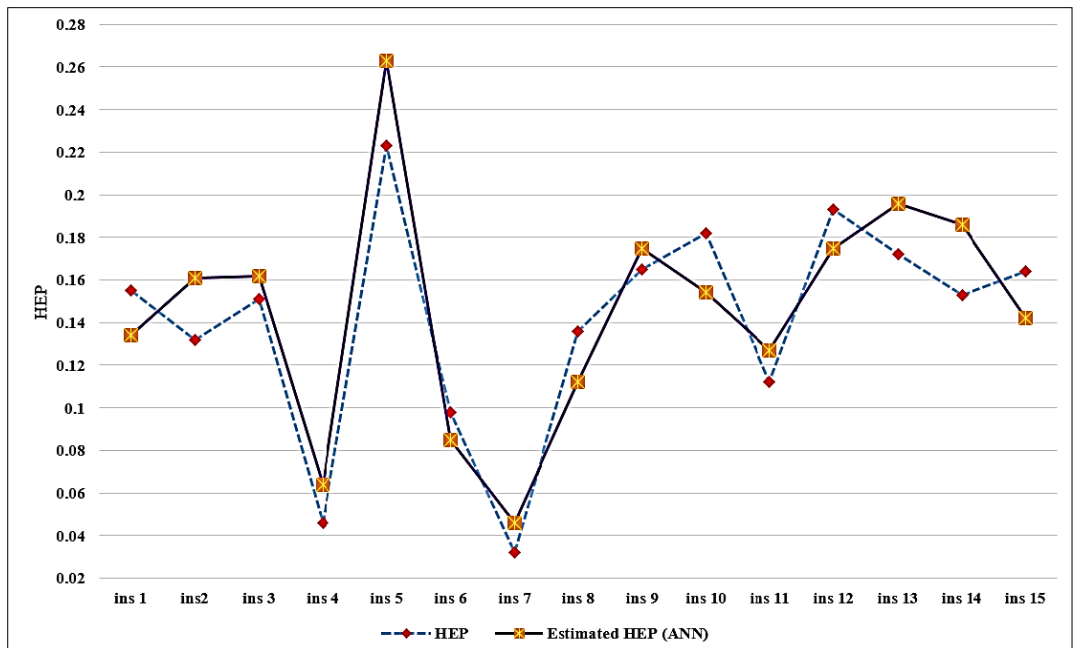

**Fig. 4. The original HEP vs. HEP proposed By ANN.**

In RSM, we have eight factors and the full RSM requires 282 runs with 10 center points. To reduce the run number we use the "Min Run Res V" method. In this method the number of runs is equal to 60 and the number of the center points is 6. *Table 4* shows the RSM design for 8 PSFs, the reliability (1-HEP) for each design has been calculated by ANN.

Table 4. The RSM design for 8 PSFs.

| Std | Run | Available Time | Stress | Complexity | Experience And Training | Procedures | Ergonomics | Fitness For Duty | Work Process | Reliability | Std | Run | Available Time | Stress | Complexity | Experience And Training | Procedures | Ergonomics | Fitness For Duty | Work Process | Reliability |
|---|---|---|---|---|---|---|---|---|---|---|---|---|---|---|---|---|---|---|---|---|---|
| 22 | 1 | 0.80 | 0.36 | 0.84 | 0.83 | 0.80 | 0.81 | 0.84 | 0.82 | 83.47 | 58 | 31 | 0.50 | 0.60 | 0.60 | 0.58 | 0.51 | 0.53 | 0.60 | 0.55 | 85.78 |
| 52 | 2 | 0.50 | 0.60 | 0.60 | 0.58 | 0.51 | 0.53 | 1.00 | 0.55 | 87.51 | 48 | 32 | 0.50 | 0.60 | 0.60 | 0.58 | 1.00 | 0.53 | 0.60 | 0.55 | 79.96 |
| 12 | 3 | 0.20 | 0.36 | 0.84 | 0.33 | 0.22 | 0.24 | 0.84 | 0.82 | 85.83 | 40 | 33 | 1.00 | 0.60 | 0.60 | 0.58 | 0.51 | 0.53 | 0.60 | 0.55 | 88.27 |
| 41 | 4 | 0.50 | 0.20 | 0.60 | 0.58 | 0.51 | 0.53 | 0.60 | 0.55 | 92.78 | 29 | 34 | 0.80 | 0.36 | 0.36 | 0.83 | 0.22 | 0.81 | 0.84 | 0.28 | 90.57 |
| 15 | 5 | 0.80 | 0.84 | 0.36 | 0.33 | 0.22 | 0.81 | 0.84 | 0.82 | 90.94 | 33 | 35 | 0.80 | 0.84 | 0.36 | 0.33 | 0.80 | 0.24 | 0.36 | 0.28 | 93.68 |
| 26 | 6 | 0.20 | 0.36 | 0.36 | 0.83 | 0.22 | 0.24 | 0.84 | 0.82 | 81.05 | 32 | 36 | 0.20 | 0.36 | 0.36 | 0.83 | 0.80 | 0.24 | 0.36 | 0.28 | 93.32 |
| 28 | 7 | 0.20 | 0.84 | 0.36 | 0.83 | 0.80 | 0.81 | 0.84 | 0.28 | 91.78 | 9 | 37 | 0.80 | 0.84 | 0.84 | 0.83 | 0.22 | 0.81 | 0.84 | 0.28 | 91.78 |
| 34 | 8 | 0.20 | 0.36 | 0.84 | 0.83 | 0.80 | 0.24 | 0.84 | 0.28 | 84.02 | 21 | 38 | 0.20 | 0.36 | 0.84 | 0.83 | 0.80 | 0.81 | 0.36 | 0.82 | 86.74 |
| 44 | 9 | 0.50 | 0.60 | 1.00 | 0.58 | 0.51 | 0.53 | 0.60 | 0.55 | 80.07 | 20 | 39 | 0.20 | 0.84 | 0.36 | 0.33 | 0.22 | 0.24 | 0.36 | 0.82 | 78.06 |
| 54 | 10 | 0.50 | 0.60 | 0.60 | 0.58 | 0.51 | 0.53 | 0.60 | 1.00 | 74.70 | 47 | 40 | 0.50 | 0.60 | 0.60 | 0.58 | 0.02 | 0.53 | 0.60 | 0.55 | 87.39 |
| 60 | 11 | 0.50 | 0.60 | 0.60 | 0.58 | 0.51 | 0.53 | 0.60 | 0.55 | 82.52 | 46 | 41 | 0.50 | 0.60 | 0.60 | 1.00 | 0.51 | 0.53 | 0.60 | 0.55 | 96.05 |
| 25 | 12 | 0.20 | 0.36 | 0.36 | 0.33 | 0.80 | 0.81 | 0.84 | 0.28 | 82.35 | 19 | 42 | 0.80 | 0.36 | 0.84 | 0.33 | 0.22 | 0.81 | 0.84 | 0.28 | 89.59 |
| 53 | 13 | 0.50 | 0.60 | 0.60 | 0.58 | 0.51 | 0.53 | 0.60 | 0.10 | 97.52 | 39 | 43 | 0.00 | 0.60 | 0.60 | 0.58 | 0.51 | 0.53 | 0.60 | 0.55 | 89.90 |
| 42 | 14 | 0.50 | 1.00 | 0.60 | 0.58 | 0.51 | 0.53 | 0.60 | 0.55 | 85.56 | 10 | 44 | 0.20 | 0.84 | 0.84 | 0.83 | 0.22 | 0.24 | 0.84 | 0.82 | 81.32 |
| 35 | 15 | 0.20 | 0.84 | 0.36 | 0.33 | 0.22 | 0.24 | 0.84 | 0.28 | 81.77 | 2 | 45 | 0.20 | 0.84 | 0.36 | 0.33 | 0.22 | 0.81 | 0.36 | 0.28 | 89.34 |
| 23 | 16 | 0.20 | 0.36 | 0.84 | 0.83 | 0.22 | 0.24 | 0.36 | 0.28 | 78.02 | 1 | 46 | 0.20 | 0.84 | 0.36 | 0.83 | 0.22 | 0.81 | 0.36 | 0.82 | 95.72 |
| 24 | 17 | 0.80 | 0.84 | 0.36 | 0.83 | 0.22 | 0.24 | 0.36 | 0.28 | 94.53 | 59 | 47 | 0.50 | 0.60 | 0.60 | 0.58 | 0.51 | 0.53 | 0.60 | 0.55 | 79.17 |
| 55 | 18 | 0.50 | 0.60 | 0.60 | 0.58 | 0.51 | 0.53 | 0.60 | 0.55 | 80.57 | 43 | 48 | 0.50 | 0.60 | 0.20 | 0.58 | 0.51 | 0.53 | 0.60 | 0.55 | 75.82 |
| 4 | 19 | 0.80 | 0.84 | 0.36 | 0.33 | 0.80 | 0.81 | 0.84 | 0.28 | 90.42 | 18 | 49 | 0.20 | 0.36 | 0.36 | 0.33 | 0.22 | 0.81 | 0.36 | 0.82 | 81.58 |
| 37 | 20 | 0.20 | 0.84 | 0.36 | 0.33 | 0.80 | 0.81 | 0.84 | 0.82 | 76.01 | 6 | 50 | 0.20 | 0.36 | 0.84 | 0.33 | 0.80 | 0.24 | 0.36 | 0.28 | 86.68 |
| 30 | 21 | 0.80 | 0.36 | 0.84 | 0.83 | 0.22 | 0.24 | 0.84 | 0.82 | 87.93 | 36 | 51 | 0.80 | 0.36 | 0.84 | 0.83 | 0.80 | 0.81 | 0.36 | 0.28 | 85.97 |
| 51 | 22 | 0.50 | 0.60 | 0.60 | 0.58 | 0.51 | 0.53 | 0.20 | 0.55 | 90.27 | 49 | 52 | 0.50 | 0.60 | 0.60 | 0.58 | 0.51 | 0.05 | 0.60 | 0.55 | 91.56 |
| 14 | 23 | 0.80 | 0.36 | 0.36 | 0.33 | 0.80 | 0.81 | 0.36 | 0.82 | 83.19 | 45 | 53 | 0.50 | 0.60 | 0.60 | 0.16 | 0.51 | 0.53 | 0.60 | 0.55 | 97.51 |
| 31 | 24 | 0.20 | 0.84 | 0.84 | 0.33 | 0.80 | 0.81 | 0.84 | 0.28 | 86.19 | 7 | 54 | 0.80 | 0.84 | 0.36 | 0.83 | 0.80 | 0.24 | 0.84 | 0.82 | 75.95 |
| 16 | 25 | 0.80 | 0.84 | 0.84 | 0.33 | 0.80 | 0.24 | 0.84 | 0.82 | 84.82 | 3 | 55 | 0.80 | 0.84 | 0.84 | 0.83 | 0.80 | 0.81 | 0.36 | 0.82 | 88.15 |
| 38 | 26 | 0.80 | 0.36 | 0.36 | 0.33 | 0.22 | 0.24 | 0.36 | 0.28 | 96.85 | 27 | 56 | 0.80 | 0.36 | 0.36 | 0.33 | 0.80 | 0.24 | 0.84 | 0.82 | 96.28 |
| 11 | 27 | 0.80 | 0.36 | 0.84 | 0.83 | 0.80 | 0.24 | 0.36 | 0.82 | 83.14 | 50 | 57 | 0.50 | 0.60 | 0.60 | 0.58 | 0.51 | 1.00 | 0.60 | 0.55 | 74.55 |
| 13 | 28 | 0.80 | 0.84 | 0.84 | 0.33 | 0.22 | 0.24 | 0.36 | 0.28 | 96.19 | 57 | 58 | 0.50 | 0.60 | 0.60 | 0.58 | 0.51 | 0.53 | 0.60 | 0.55 | 92.67 |
| 56 | 29 | 0.50 | 0.60 | 0.60 | 0.58 | 0.51 | 0.53 | 0.60 | 0.55 | 96.37 | 5 | 59 | 0.20 | 0.84 | 0.84 | 0.83 | 0.80 | 0.24 | 0.36 | 0.28 | 94.98 |
| 17 | 30 | 0.80 | 0.36 | 0.84 | 0.83 | 0.22 | 0.81 | 0.36 | 0.82 | 78.21 | 8 | 60 | 0.20 | 0.84 | 0.84 | 0.33 | 0.22 | 0.81 | 0.36 | 0.82 | 76.41 |

The above designs have been investigated by RSM, and a quadratic model has been selected to use for the fitting process. The result of the ANOVA test is shown in *Table 5*. According to *Table 5*, the Model F-value of 4.65 implies the model is significant. There is only a 0.01% chance that a "Model F-Value" this large could occur due to noise. Also, the "Lack of Fit F-value" of 0.38 implies the "Lack of Fit" is not significant relative to the pure error. There is a 96.30% chance that a "Lack of Fit F-value" this large could occur due to noise.

**Table 5. The result of ANOVA for RSM designs.**

| Source | Sum Of Squares | Df | Mean Square | F Value | P-Value | |
|---|---|---|---|---|---|---|
| **Model** | 8.0622E+11 | 15 | 5.37E+10 | 4.64885 | < 0.0001 | Significant |
| **A-Available Time** | 60659292101 | 1 | 6.07E+10 | 5.246629 | 0.0268 | |
| **B-Stress** | 163330712.9 | 1 | 1.63E+08 | 0.014127 | 0.9059 | |
| **C-Complexity** | 2855478876 | 1 | 2.86E+09 | 0.24698 | 0.6217 | |
| **D-Experience And Training** | 1238037384 | 1 | 1.24E+09 | 0.107082 | 0.7450 | |
| **F-Ergonomics** | 3675806671 | 1 | 3.68E+09 | 0.317933 | 0.5757 | |
| **G-Fitness For Duty** | 9371607097 | 1 | 9.37E+09 | 0.810582 | 0.3728 | |
| **H-Work Process** | 2.08453E+11 | 1 | 2.08E+11 | 18.0298 | 0.0001 | |
| **AD** | 92996049608 | 1 | 9.3E+10 | 8.043546 | 0.0069 | |
| **AF** | 23260353958 | 1 | 2.33E+10 | 2.011867 | 0.1631 | |
| **BD** | 36896245096 | 1 | 3.69E+10 | 3.191282 | 0.0809 | |
| **BF** | 41169334214 | 1 | 4.12E+10 | 3.560876 | 0.0658 | |
| **BG** | 41678085594 | 1 | 4.17E+10 | 3.60488 | 0.0642 | |
| **DF** | 47101313124 | 1 | 4.71E+10 | 4.073954 | 0.0497 | |
| **C^2** | 95926695605 | 1 | 9.59E+10 | 8.297028 | 0.0061 | |
| **D^2** | 1.4491E+11 | 1 | 1.45E+11 | 12.53377 | 0.0010 | |
| **Residual** | 5.08709E+11 | 44 | 1.16E+10 | | | |
| **Lack of Fit** | 3.79594E+11 | 39 | 9.73E+09 | 0.376917 | 0.9630 | Not significant |
| **Pure Error** | 1.29116E+11 | 5 | 2.58E+10 | | | |
| **Cor Total** | 1.31493E+12 | 59 | | | | |

The final equation proposed by RSM is as follows:

Reliability^3 = +8.21001E+005 +7.00605E+005*Available Time -2.14049E+005 *Stress +1.41735E+006*Complexity -1.79156E+006* Experience And Training -4.82962E+005 *Ergonomics +2.92725E+005 * Fitness For Duty -2.62798E+005 * Work Process -6.94343E+005 * Available Time * Experience And Training -3.17430E+005 * Available Time * Ergonomics +5.35751E+005 * Stress * Experience And Training +5.10007E+005 * Stress * Ergonomics -5.93368E+005 * Stress * Fitness For Duty +5.21145E+005 * Experience And Training * Ergonomics -1.21129E+006 * Complexity^2 +1.35036E+006 * Experience And Training^2.

To illustrate the robustness of the proposed equation by RSM the original reliability and the predicted reliability are shown in *Fig. 5*. The proposed graph indicates that the RSM equation effectively predicts the reliability value for each design.

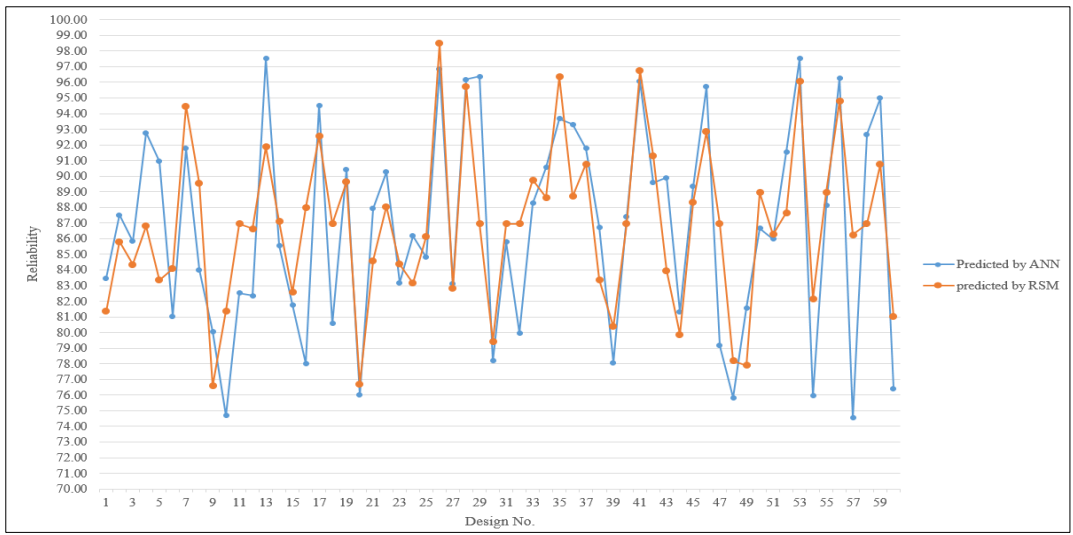

**Fig. 5. The original reliability and the predicted reliability.**

Other graphs such as residual normal plot, and residual vs. predicted are shown in *Fig. 6*. *Fig. 6.a* shows that all points fall on the line and the framework fits the data well. For a well-fitted model, the residuals plotted with respect to predicted values should not follow any particular pattern and should be symmetrically distributed with respect to the center line, *Fig. 6b* follows these rules and indicates that the framework is well fitted.

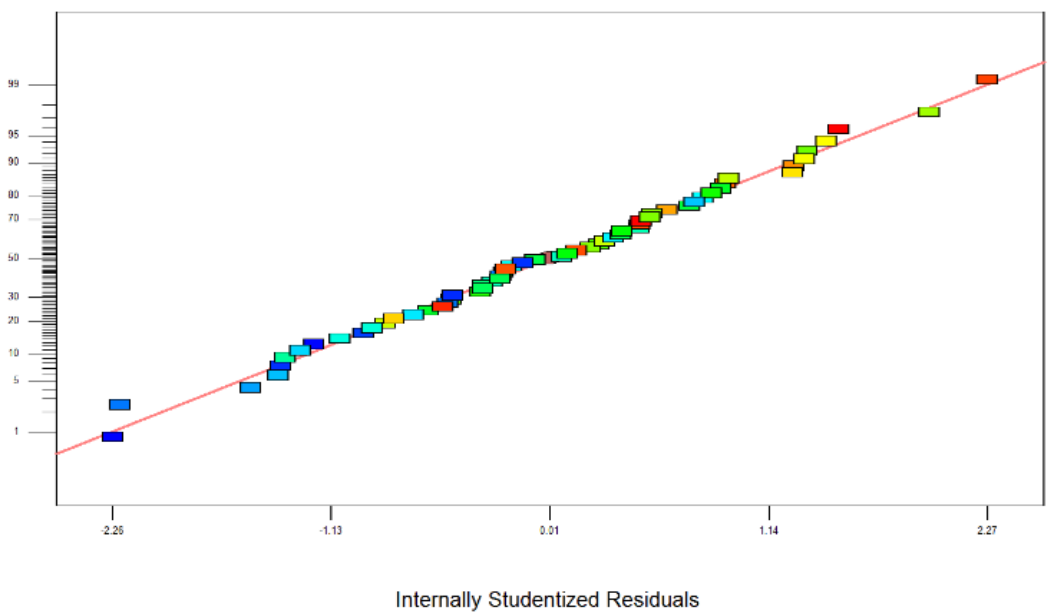

a

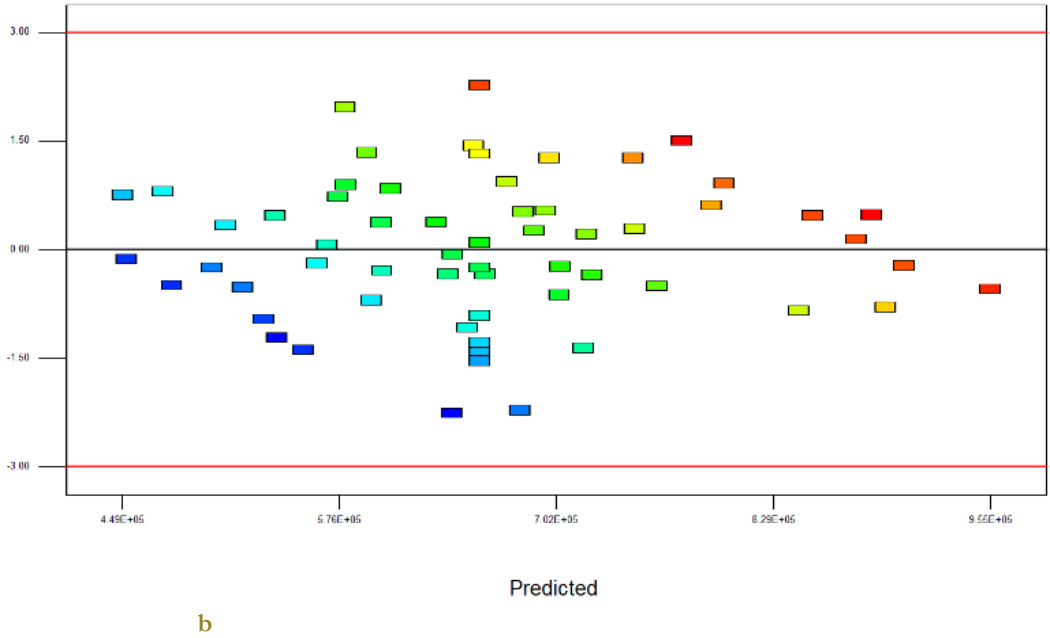

b

**Fig. 6. The related graph for RSM: a. normal plot of residuals, b. residuals vs. predicted.**

It is understood from the obtained results that the fourth PSF (Procedures) is not effective in reliability (1-HEP) and we can eliminate this PSF to increase the accuracy of the ANN. The RSM can assess the importance of each PSF and the ANN calculates the HEP considering the PSFs dependency. In the next step, the ANN is utilized again to estimate the HEP for the proposed instances in *Table 1*. The "procedure" is eliminated from non-effective PSF and the proposed ANN is run with 7 PSFs as Inputs. Other parameters of ANN do not change. The provided results are shown in *Table 6*.

**Table 6. The result of ANN after RSM method.**

|  | HEP | Estimated HEP-Before RSM | Estimated HEP-After RSM | SE-Before RSM | SE-After RSM |
|---|---|---|---|---|---|
| **Ins 1** | 0.155 | 0.134 | 0.141409178 | 4.41E-04 | 1.85E-04 |
| **Ins2** | 0.132 | 0.161 | 0.149928035 | 8.41E-04 | 3.21E-04 |
| **Ins 3** | 0.151 | 0.162 | 0.158017965 | 1.21E-04 | 4.93E-05 |
| **Ins 4** | 0.046 | 0.064 | 0.055952861 | 3.24E-04 | 9.91E-05 |
| **Ins 5** | 0.223 | 0.263 | 0.250007036 | 1.60E-03 | 7.29E-04 |
| **Ins 6** | 0.098 | 0.0851 | 0.089366754 | 1.66E-04 | 7.45E-05 |
| **Ins 7** | 0.032 | 0.046 | 0.042212003 | 1.96E-04 | 1.04E-04 |
| **Ins 8** | 0.136 | 0.112 | 0.11851672 | 5.76E-04 | 3.06E-04 |
| **Ins 9** | 0.165 | 0.175 | 0.170615192 | 1.00E-04 | 3.15E-05 |
| **Ins 10** | 0.182 | 0.154 | 0.161680555 | 7.84E-04 | 4.13E-04 |
| **Ins 11** | 0.112 | 0.127 | 0.121836516 | 2.25E-04 | 9.68E-05 |
| **Ins 12** | 0.193 | 0.175 | 0.183575021 | 3.24E-04 | 8.88E-05 |
| **Ins 13** | 0.172 | 0.196 | 0.186279252 | 5.76E-04 | 2.04E-04 |
| **Ins 14** | 0.153 | 0.186 | 0.17487524 | 1.09E-03 | 4.79E-04 |
| **Ins 15** | 0.164 | 0.142 | 0.152283547 | 4.84E-04 | 1.37E-04 |

The SE-After RSM value indicates that the accuracy of proposed ANN increases in cooperation with RSM. The decreases in the SE are shown in *Fig. 7*. The tolerance between original HEP and predicted HEP has decreased in comparison with ANN without RSM.

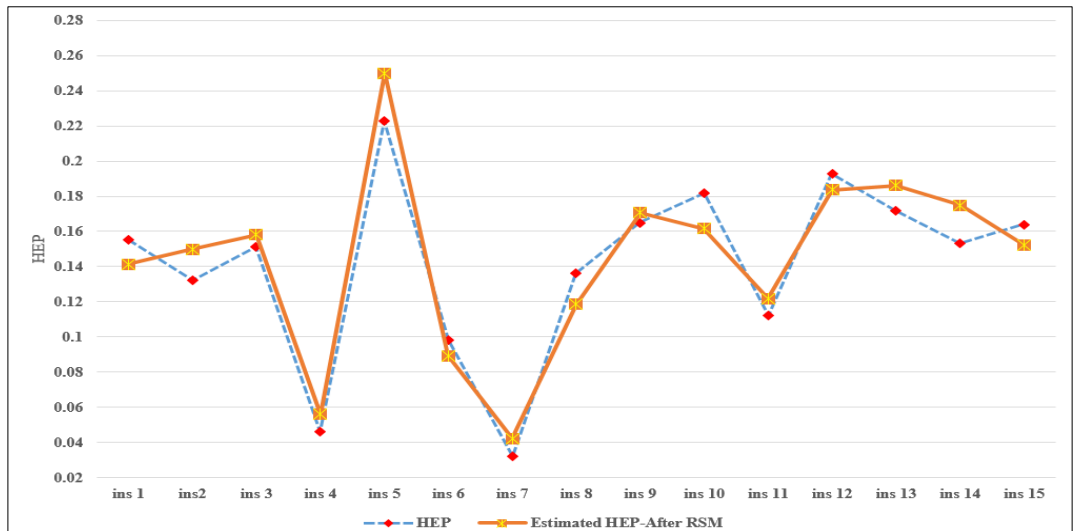

**Fig. 7. The original HEP vs. HEP proposed By ANN after RSM implementation**.

## 4 | Conclusion

HRA is an important issue in production systems and some methods should be proposed to evaluate the reliability of humans in these systems. All HRA methods proposed some Performance Shaping Factors (PSF) and calculate the HEP base on these factors. Selecting the most effective PSFs and considering the PSFs dependency is an important problem that decreases the time and cost and increases the HRA accuracy. In this paper, we proposed an ANN based Human Reliability Analysis (ANNHRA) in cooperation with RSM, with ANN we can calculate the HEP considering the PSFs dependency and using RSM we can find the most effective PSFs. This framework can provide more accurate HEP with lower time and cost. The performance of the proposed framework was examined and the provided results indicated the model can obtain efficient and effective HEP and reliability value in production systems.